\documentclass{iopjournal}

\usepackage{amsmath}
\usepackage{xcolor}

\begin{document}

\articletype{Article type} %	 e.g. Paper, Letter, Topical Review...

\title{Material-Specific Mapping of Plasmonic Modal Dispersion via Discrete Momentum-Space Probes}

\author{Youssef El Badri$^{1,*}$ \orcid{0000-0003-4682-4446}, Hicham Mangach$^2$ \orcid{orcid.org/0000-0003-1056-2608}, Yan Pennec$^2$ \orcid{orcid.org/0000-0002-4719-9913}, Bahram Djafari-Rouhani$^2$\orcid{orcid.org/0000-0001-6983-9689},\\
Abdenbi Bouzid$^1$, Mustapha Bahich$^1$, Younes Achaoui$^1$}

\affil{$^1$ Laboratory of Optics, Information Processing, Mechanics, Energetics and Electronics, Department of Physics, Moulay Ismail University, B.P. 11201, Zitoune, Meknes, Morocco.}

\affil{$^2$ Institut d’Electronique, de Microélectonique et de Nanotechnologie, UMR CNRS 8520, Département de Physique, Université de Lille, 59650 Villeneuve d’Ascq, France.}

\affil{$^*$Youssef El Badri.}

\email{y.elbadri@edu.umi.ac.ma}

\begin{abstract}
Accurate measurement of surface plasmon polariton (SPP) dispersion remains challenging, as conventional angle-resolved techniques are limited by surface quality, diffraction artifacts, and geometry-induced band folding. Here, we show that SPP dispersion can be reconstructed from transmission spectra of plasmonic gratings with subwavelength apertures acting as Fabry--P\'erot (FP) cavities. The approach harnesses modal hybridization between localized FP modes and SPPs, resolved using non-Hermitian eigenmode decomposition and validated by finite-difference time-domain simulations. $\omega$--$k$ dispersion mapping is achieved by varying the grating periodicity, with each structure probing a distinct in-plane momentum state. Geometry- and material-dependent corrections for aperture-induced leakage and dispersive phase shifts yield reconstructed relations in close agreement with eigenmode calculations across non-dispersive, Drude, and Drude--Lorentz models as well as experimental optical datasets spanning metals, oxides, and nitrides. The method is material-agnostic and requires no momentum-resolved instrumentation. Sensitivity to fabrication-induced wall roughness is also assessed: FP resonance positions remain spectrally stable with no measurable linewidth broadening across the explored perturbation range, and the modal field topology is largely preserved throughout. However, transmitted amplitude decreases monotonically owing to enhanced ohmic absorption at the perturbed boundaries.
\end{abstract}

\section{Introduction} \label{sec1}
The rapid development of nano-plasmonics has introduced persistent challenges in the accurate characterization of plasmonic materials, arising from both their complex optical response and the limitations of existing metrology techniques \cite{baburin2019silver,foteinopoulou2019feature}. At optical frequencies, plasmonic materials exhibit strong dispersion, high reflectivity, anisotropy, and nonlinear effects, while remaining highly sensitive to oxidation, surface contamination, and microstructural variations \cite{oates2013effect,jiang2016grain,gorkunov2014tarnishing}. Because plasmonic resonances depend nonlinearly on the dielectric function, even modest changes in plasma frequency or damping can produce substantial spectral shifts and linewidth broadening \cite{wu2014intrinsic}. For example, the dipolar plasmon resonance condition for a metallic nanoparticle in a homogeneous environment is $\Re\{\varepsilon(\omega^*)\}\approx -2\,\varepsilon_{\rm env}$, where $\omega^*$ denotes the resonance frequency. This relation implies a resonance sensitivity proportional to $1/|\partial\varepsilon'/\partial\omega|$, so that the spectral response becomes especially pronounced in frequency regions where $\varepsilon'$ varies slowly with frequency \cite{miller2005sensitivity}. Near inter-band transitions, $\varepsilon'$ can also be strongly affected by microstructural imperfections and defects, further amplifying sample-to-sample variability \cite{derkachova2016}. In practice, variations in fabrication and synthesis protocols, including differences in grain morphology, deposition conditions, substrate adhesion layers, and residual contamination, continue to produce significant discrepancies in reported dielectric functions, even for nominally high-purity plasmonic films \cite{mcpeak2015plasmonic,jiang2016realistic,coviello2024accurate}. Although careful spectroscopic measurements have yielded increasingly refined tabulated datasets \cite{mcpeak2015plasmonic,yang2015optical,babar2015optical,jiang2016realistic}, persistent disagreements among them indicate that nominal material purity alone is insufficient to ensure a universally transferable dielectric function \cite{baburin2019silver,coviello2024accurate,lebsir2022ultimate,al2022t,casses2022quantitative}. For example, Lebsir \textit{et al.} demonstrated that selecting among published dielectric function datasets for gold, which is a well-studied noble metal with comparatively mature optical constant libraries, produces variations of approximately $50\%$ in predicted SPP propagation lengths for nominally identical samples \cite{lebsir2022ultimate}. These discrepancies are not merely academic: changes in the adopted optical constants can lead to appreciable shifts in predicted resonance frequencies, quality factors, and near-field enhancement levels, thereby compromising reproducibility and predictive design \cite{christopoulos2019calculation,coviello2024accurate,habteyes2012metallic}. Thus, high-precision plasmonic engineering requires sample-specific or in-situ determination of optical properties rather than exclusive reliance on tabulated literature values \cite{colin2020situ}.\\
A direct route to characterizing plasmonic behavior is through dispersion diagrams that map optical frequency to in-plane momentum, encoding propagation constants, group velocity, modal confinement, and optical loss \cite{de2025accessing,barbara2010plasmon}. For metal-dielectric interfaces, the surface plasmon polariton (SPP) dispersion is governed by the material permittivity and the electromagnetic boundary conditions, making the dispersion relation a sensitive probe of the underlying plasmonic response. Conventional retrieval techniques, such as prism-coupled angle-resolved configurations in Kretschmann and Otto geometries, identify SPP resonances from reflectivity minima measured as a function of incidence angle or wavelength \cite{maier2007plasmonics,das2023computational,ohad2018spatially}. While highly effective for flat films, these methods require angular scanning and precise optical alignment. Moreover, the retrieved dispersion reflects not only the plasmonic response of the material but also the prism configuration, substrate characteristics, and dielectric environment \cite{butt2025surface,yesudasu2021recent,thadson2022measurement,nishida2020evaluation}. Surface roughness, film inhomogeneity, finite-thickness effects, and refractive-index fluctuations can further broaden the measured resonances, complicating interpretation \cite{raether1988surface,hojjat2025surface, treebupa23sens}. Periodic plasmonic arrays offer an alternative by exciting Bloch SPP modes through reciprocal lattice vectors that compensate for the photon-plasmon momentum mismatch, with resonances appearing as absorption minima \cite{raether2006surface,ghaemi1998surface}. However, this mechanism folds the continuous flat-interface SPP branch into the first Brillouin zone, yielding geometry-dependent Bloch bands whose spectral response is strongly shaped by lattice periodicity \cite{de2025accessing}. Furthermore, Wood's anomalies introduce spectral perturbations that can overlap with or distort the underlying plasmonic resonances \cite{ghaemi1998surface,gao2008screening,Zhou2010}.\\
Resonant photonic architectures provide an alternative strategy for probing plasmonic behavior through cavity-mediated interactions. In particular, subwavelength Fabry--P\'erot (FP) resonators embedded within plasmonic gratings support strongly confined electromagnetic modes inside narrow apertures, enhancing light-matter interaction within the cavity volume \cite{d2011Aguanno}. FP cavity modes can hybridize with SPPs at the metal-dielectric interface, producing spectrally sharp resonances with strong sensitivity to both geometric and material parameters \cite{d2011Aguanno,marquier2005resonant}. Unlike conventional grating-coupled SPPs, where the field is distributed primarily along the interface, subwavelength apertures sustain localized waveguide modes that form FP resonances within the slit volume, so that the slit acts as an active resonant cavity whose eigenfrequencies are directly modified by plasmonic coupling \cite{d2011Aguanno,dechaux2016influence,jouy2011coupling}. Since resonance positions arise from FP-SPP hybridization, variations in plasmonic coupling manifest directly as eigenfrequency shifts rather than being inferred indirectly from reflectivity minima \cite{dechaux2016influence,d2011Aguanno}. The localized interaction also enhances sensitivity to local optical variations beyond what is accessible through spatially averaged interface responses \cite{baumberg2019extreme}. Recent experiments have demonstrated nearly two orders of magnitude enhancement in terahertz detector responsivity using FP-coupled plasmonic resonators \cite{muravev2025plasmonic}, highlighting the strong field localization achievable in such hybrid systems. Collectively, these properties make FP-SPP resonances attractive probes for spectrally resolved reconstruction of plasmonic dispersion relations.\\
In this work, we numerically demonstrate an approach for probing plasmonic dispersion based on extraordinary optical transmission through subwavelength gratings. By systematically varying the grating periodicity, each structure samples a discrete in-plane momentum state, with the spectral positions of FP--SPP resonances encoding the corresponding dispersion information, enabling mapping without angular scanning or momentum-resolved instrumentation. In Section 3, we establish the modal landscape of the plasmonic grating as a function of periodicity using a quasi-normal mode (QNM) formulation cross-validated with finite-difference time-domain (FDTD) simulations. In Section 4, we quantify pure geometric effects in a perfect electric conductor (PEC) and determine the deviation from an ideal FP cavity via a correction factor $\sigma(r)$. Section 5 reintroduces material dispersion and examines the influence of permittivity on the transmission response. In Section 6, we demonstrate a numerical proof-of-concept for plasmonic dispersion mapping, tracking FP--SPP resonance shifts across discrete momentum states to reconstruct flat-interface SPP dispersion using a composite correction factor $\sigma(r,\varepsilon)$. The method is validated across representative plasmonic materials and dispersion scenarios. Finally, Section 7 examines sensitivity to fabrication-induced wall roughness: resonance positions remain spectrally stable across the explored perturbation range, with no significant linewidth broadening and only minor eigenfrequency shifts within the $\sigma(r,\varepsilon)$ correction window, while roughness primarily reduces transmitted through increased ohmic absorption.
\section{Materials and Methods} \label{sec2}
\begin{figure}
\centering
\includegraphics[width=1\linewidth]{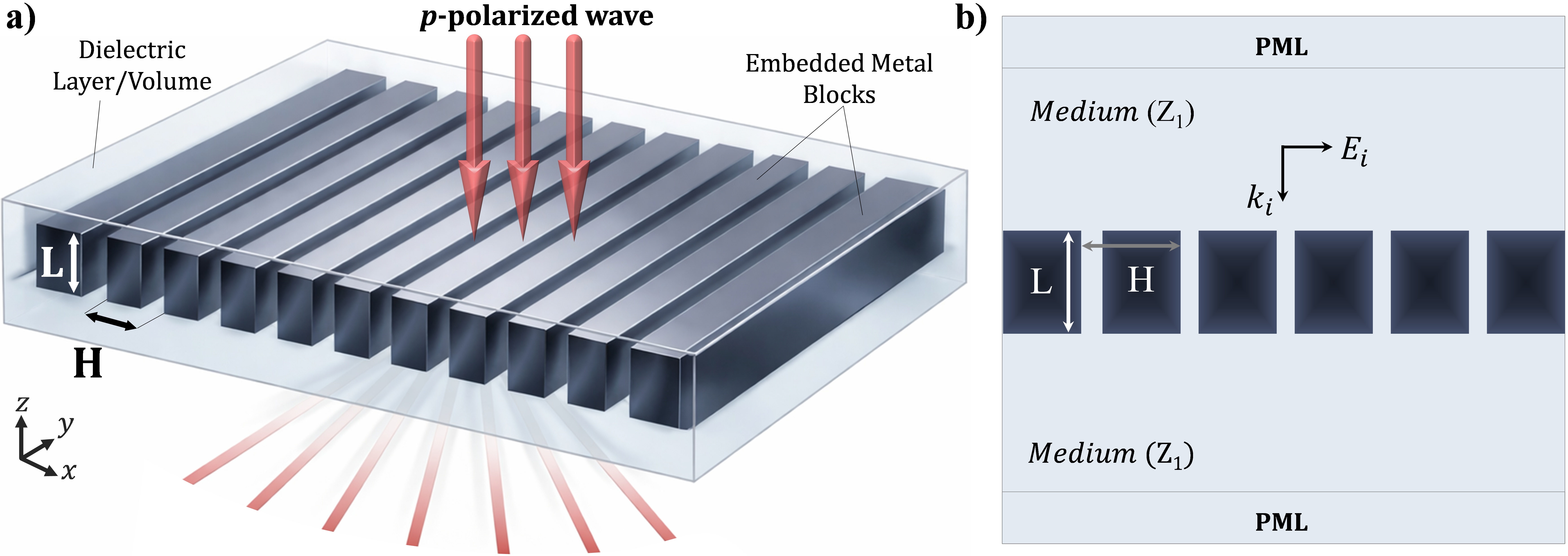}
\caption{\textbf{a)} Three-dimensional rendering and \textbf{b)} schematic representation of the numerical model. The structure consists of a plasmonic grating with subwavelength apertures embedded in a dielectric matrix where the substrate, superstrate and slit filling are composed of the same dielectric material. H and L denote the horizontal (periodic) and lateral (thickness) parameters of the unit cell.}
\label{fig:Figure 1}
\end{figure}

The system under investigation consists of an optically opaque metallic slab perforated by periodic subwavelength apertures, with a prescribed filling factor defining the aperture size (Fig~\ref{fig:Figure 1}). The metallic blocks are embedded in the same dielectric on both sides, i.e., identical superstrate, substrate and slit filling, eliminating impedance mismatch at the interfaces. This symmetric embedding ensures that the reflection coefficient at each aperture entrance is governed solely by the metal-dielectric impedance contrast, which is the calibrated quantity underlying the FP resonance analysis. Any index mismatch between the slit fill and the surrounding half-spaces would modify the end-reflection coefficients, shifting the FP resonance frequencies away from this calibrated condition and introducing a systematic error in the recovered SPP dispersion. The lateral and horizontal dimensions of the grating are kept equal ($H=L$) for simplicity of scaling analysis. Owing to translational symmetry along the slit direction ($y$-axis), the analysis is restricted to the $x$-$z$ plane, where the electric field has components $E_x$ and $E_z$ and the magnetic field is oriented along the $y$-axis. Under normal incidence, only transverse magnetic (TM) modes couple to the slit apertures, while TE modes are suppressed \cite{garcia2010light}. Thus, all simulations employ TM-polarized illumination. The optical response is modeled using complementary computational electromagnetic techniques: the Finite Element Method (FEM) in COMSOL Multiphysics and the Finite-Difference Time-Domain (FDTD) method in ANSYS Lumerical. FEM provides frequency-domain modal analysis, while FDTD captures time-domain spectral behavior, allowing cross-validation of results. The simulation domain comprises one unit cell of the grating with periodic (Floquet–Bloch) boundary conditions laterally and perfectly matched layers (PMLs) at the top and bottom to absorb outgoing waves. Initially, the metal is modeled as a PEC to isolate geometric effects. Afterward, we explore a dispersive case described using Drude permittivity model: $\varepsilon(\omega)=\varepsilon_\infty-\omega_p^2/[\omega^2 + i \gamma \omega]$. With the background constant $\varepsilon_\infty=1$, damping coefficient $\gamma=1.3\times10^{13}$ rad/s, and plasma frequency $\omega _p =3.1\times10^{16}$ rad/s.\\
Modal decomposition provides physical insight into the contribution of individual photonic modes to the overall electromagnetic response. Since plasmonic materials are both lossy and dispersive, the associated eigenproblem is inherently non-Hermitian. In such systems, the Maxwell propagation operator becomes non-self-adjoint and the Helmholtz equation forms a nonlinear eigenvalue problem. Previous work has shown that by introducing auxiliary polarization and current density fields $(\mathbf{P},\mathbf{J})$, the dispersive eigenproblem can be linearized \cite{raman2010photonic}. The resulting formulation is implemented within a weak FEM framework and solved using the COMSOL eigenmode solver \cite{yan2018rigorous}.
\begin{equation}\label{eq:01}
\left[
\begin{array}{cccc}
0 & -i \mu_0^{-1} \nabla \times & 0 & 0 \\
i \varepsilon_\infty^{-1} \nabla \times & 0 & 0 & -i \varepsilon_\infty^{-1} \\
0 & 0 & i \omega_p^2 \varepsilon_\infty & i \\
0 & 0 & i \omega_0^2& -i \gamma
\end{array}
\right]
\left[
\begin{array}{c}
\mathbf{H}(\mathbf{r}, \omega) \\
\mathbf{E}(\mathbf{r}, \omega) \\
\mathbf{P}(\mathbf{r}, \omega) \\
\mathbf{J}(\mathbf{r}, \omega)
\end{array}
\right]
=
\left( \frac{\omega}{c} \right)^2
\left[
\begin{array}{c}
\mathbf{H}(\mathbf{r}, \omega) \\
\mathbf{E}(\mathbf{r}, \omega) \\
\mathbf{P}(\mathbf{r}, \omega) \\
\mathbf{J}(\mathbf{r}, \omega)
\end{array}
\right]
\end{equation}
In parallel, FDTD simulations are performed by injecting a broadband pulse from multiple randomly distributed point sources within the unit cell to ensure that all eigenmodes are sufficiently excited. Identification of Bloch mode frequencies is done by computing the power spectral density (PSD) via a Fast Fourier Transform of the steady-state fields collected by spatial monitors, thereby extracting the resonant frequencies and associated field distributions \cite{chan1995order}. This approach can straightforwardly deal with the optical dispersion of metals through the discretization of the constitutive relations that link the electric displacement vector to the electric field.
The photonic dispersion diagrams are constructed along the high-symmetry path $\Gamma X$ in the irreducible Brillouin zone (IBZ), ranging from $k_x=0$ to $k_x=\pi/H$. To further characterize mode behavior, we introduce the energy localization ratio, denoted by $\Theta$, which quantifies the spatial distribution of electromagnetic energy across different regions of the structure. $\Theta$ is defined as follows: 
\begin{equation}\label{eq:02}
\Theta = \frac{\iiint_{v_{interest}} \textbf{w}_{en}(r) \, dV}{\iiint_{v_{\text{total}}} \textbf{w}_{en}(r) \, dV}
\end{equation}
where $\textbf{w}_{en}(r)$ denotes the time-averaged electromagnetic energy density. The numerator represents the energy stored within a specified region of interest (e.g., the cavity, metal, or surrounding dielectric), while the denominator corresponds to the total electromagnetic energy stored in the structure. In our case, the targeted domain is taken to be the metallic blocks. This metric enables the classification of modes according to their dominant energy localization and provides a direct visual correlation between modal frequency and spatial field concentration. In the dispersion diagrams, $\Theta$ is overlaid as a colormap to highlight regions of strong modal localization.\\
To assess the sensitivity of the FP-based transmission mapping to fabrication imperfections, stochastic roughness was introduced along the metallic boundaries of the subwavelength apertures using a power-law Fourier synthesis implemented via parametric curve nodes in COMSOL Multiphysics. The wall distortion was represented by a truncated trigonometric series of the form:
\begin{equation}\label{eq:03}
\delta(s)=\Omega\;\sum_{n=1}^{N}\frac{g(n)}{n^{\alpha}}
\cos\left(2\pi n s+f(n)\right),
\end{equation}
where $s\in[0,1]$ is the normalized arc-length coordinate, $g(n)$ are independent zero-mean unit-variance Gaussian random coefficients, $f(n)$ are uniformly distributed random phases in $[0,2\pi]$, $N$ is the spectral cutoff order, and $\alpha$ is the spectral decay exponent. Small values of $\alpha$ give greater relative weight to high-spatial-frequency components, thereby yielding rougher wall profiles. The scaling factor $\Omega$ was set to $10\%$, $15\%$, and $20\%$ of the nominal aperture width for the three configurations considered. The stochastic boundary perturbation also produces small variations in the effective aperture width and, consequently, in the filling factor. These variations are an inherent consequence of the roughness generation procedure and were retained to provide a more realistic assessment of the robustness of the proposed mapping approach. In all cases, the spectral parameters were held fixed at $N=60$ and $\alpha=1.1$, thereby preserving the same roughness spectrum while varying only the overall perturbation strength. The resulting profile was applied as a normal displacement to the nominal wall geometry, generating stochastic boundary perturbations while maintaining the nominal grating topology. For each configuration, a single realization with a fixed random seed was used to ensure reproducibility across mesh refinements and parameter sweeps. The perturbed geometries were analyzed using the same numerical settings as in the reference smooth-wall case, enabling a direct comparison of the reconstructed dispersion map.
\begin{figure}
\centering
\includegraphics[width=1\linewidth]{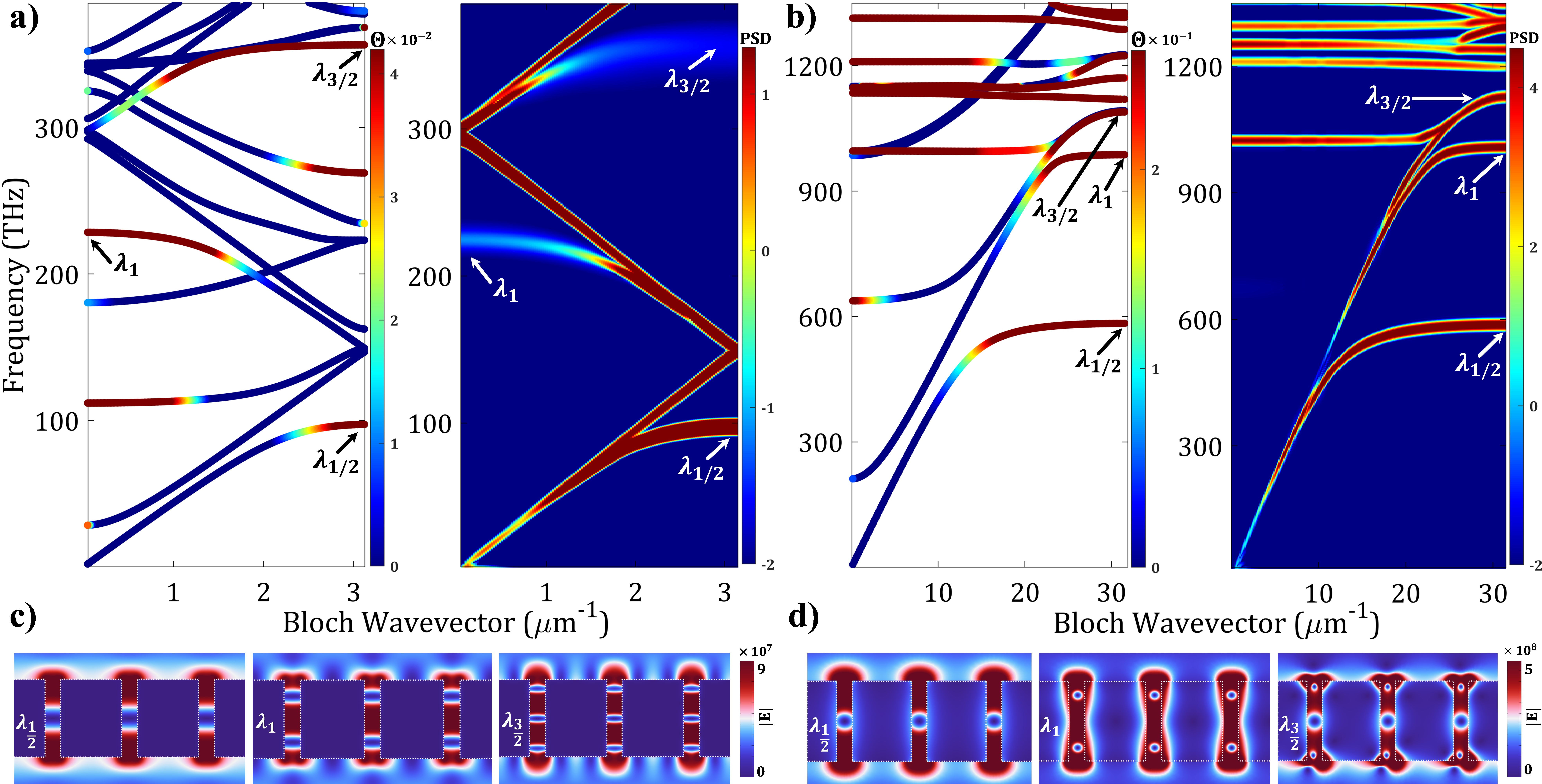}
\caption{\textbf{a)} and \textbf{b)} Photonic dispersion diagrams evaluated using QNM-FEM (left) and FDTD (right) for two sizes, $H=1$ µm and $H=100$ nm, respectively. The color palettes in these analyses represent the PSD for FDTD and $\Theta $ for FEM. \textbf{c)} and \textbf{d)} Eigenvectors depicting the modulus of the electric field at $\lambda_{1/2}$, $\lambda_1$, and $\lambda_{3/2}$ for the two sizes.}
\label{fig:Figure 2}
\end{figure}
\section{Modal Analysis in Dispersive Media}\label{sec3}
\indent Enhanced transmission through plasmonic structures is governed by two primary mechanisms. Porto \textit{et al.} showed that TM-polarized light can couple either to resonant cavity modes confined within subwavelength apertures or to surface plasmon modes propagating along metal interfaces, both capable of near-unity transmittance \cite{porto1999transmission}. Subsequent studies confirmed that the relative dominance of these two pathways depends on geometry, wavelength, and material loss \cite{garcia2010light, garcia2007colloquium}. While Marquier \textit{et al.} later described hybridization between FP and SPP modes, manifesting as anti-crossing and frequency shifts influenced by grating symmetry and material loss, the dispersion diagram remained schematic and neglected diffractive and near-field coupling at high frequencies \cite{marquier2005resonant}. The present work extends this understanding through a rigorous full-wave modal decomposition based on QNM, enabling explicit construction of dispersion diagrams and direct mapping of hybridized eigenstates, particularly in the high-frequency regime, which is substantially dispersive. The material blocks are modeled with a generic Drude-type metal. Fig~\ref{fig:Figure 2} presents dispersion diagrams of the plasmonic gratings for two periodicities to highlight the contrast in photonic behavior at different frequency spans. For each eigenvalue, the QNM analysis embeds its energy localization ratio, revealing how FP standing waves within apertures couple to surface-bound modes. Complementary FDTD simulations yield the optical energy coupling for resonance frequencies. FEM captures both radiative and evanescent states, whereas FDTD emphasizes resonant ones, since the excitation of all modes is not guaranteed, particularly near degeneracies. The excellent agreement between both approaches validates the QNM framework for plasmonic eigenstates analysis.\\
For larger periodicities, $H = 1~\mu$m, the dispersion diagrams exhibit the anti-crossing branches characteristic of the classical grating-coupled SPP resonant-cavity modes at half-wavelength multiples $(m-1)\lambda/2$, where $m$ denotes the number of field maxima \cite{marquier2005resonant}. The cavity resonance order is defined using the number of maxima in the electric field norm along the aperture, which is directly observable in the eigenvector maps. In this convention, the effective cavity length satisfies: $L_{eff}=(m-1)\lambda/2$, where $m$ denotes the number of local maxima. Fig~\ref{fig:Figure 2}.c and d display the associated eigenvector field maps depicting the electric field norm for the two sizes, respectively. These are the fundamental standing-wave patterns for $\lambda_{1/2}$ (two hot spots), $\lambda_1$ (three hot spots), and $\lambda_{3/2}$ (four hot spots). The lowest TM mode shows no cut-off frequency, forming a linear band that folds at Brillouin zone boundaries. Bloch branches are modulated by FP resonances that emerge into the evanescent zone. Beyond $\lambda_{3/2}$, higher-order diffraction modes dominate, suppressing propagation.
\begin{figure}[b]
\centering
\includegraphics[width=0.95\linewidth]{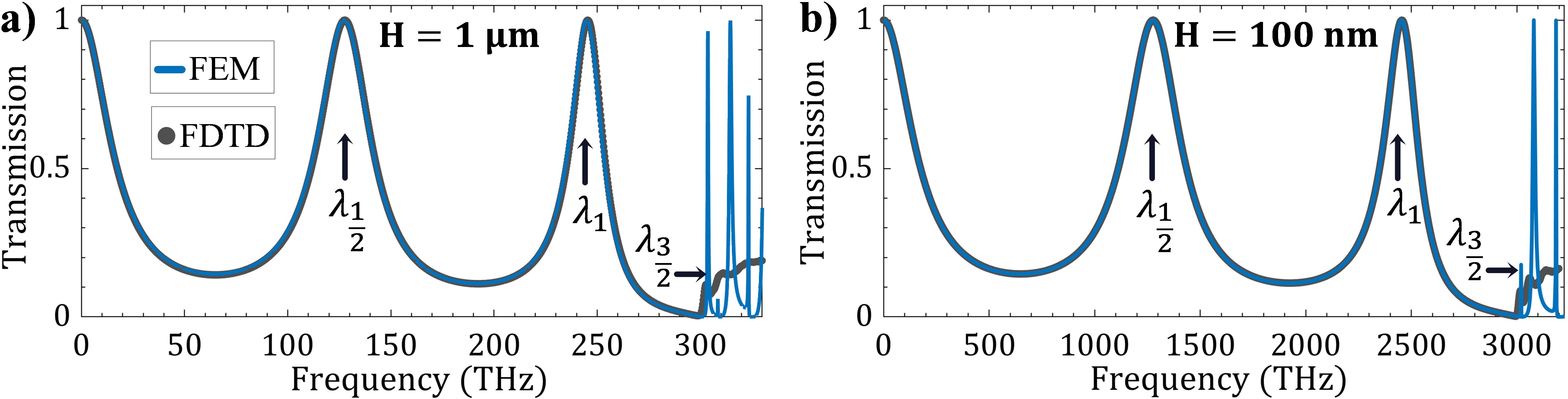}
\caption{\textbf{a)} and \textbf{b)} Transmission responses evaluated using FEM and FDTD through the PEC slab, with a filling factor of $20\%$, for two period sizes: 1 $\mu$m, $100$ nm, respectively. The FP resonant peaks $(\lambda_{1⁄2},\lambda_1,\lambda_{3/2})$ are depicted.}
\label{fig:Figure 3}
\end{figure}
Reducing the periodicity to $H = 100 nm$ drastically reshapes the modal landscape. Resonance modes enter the first evanescent zone before band folding, exhibiting flattening and anti-crossing due to the increased reciprocal lattice vector and modal suppression brought on by the material's intrinsic electronic constraints (plasmonic threshold). Strong hybridization between FP and SPP branches is confirmed by both the energy localization $\Theta$ and the associated field maps, highlighting enhanced energy penetration into the metal blocks at this frequency regime. At frequencies approaching the material plasma frequency ($\sim1160$ THz), the interface no longer supports only tightly bound SPP and the hybridization reorganizes. A complex coupling among localized plasmons, Bloch-SPP, and aperture waveguide modes produces flat bands with near-zero group velocity. These are hallmarks of strong FP-SPP photonic-plasmonic hybridization and aspects that would alter the transmission signature.
\begin{figure}
\centering
\includegraphics[width=0.98\linewidth]{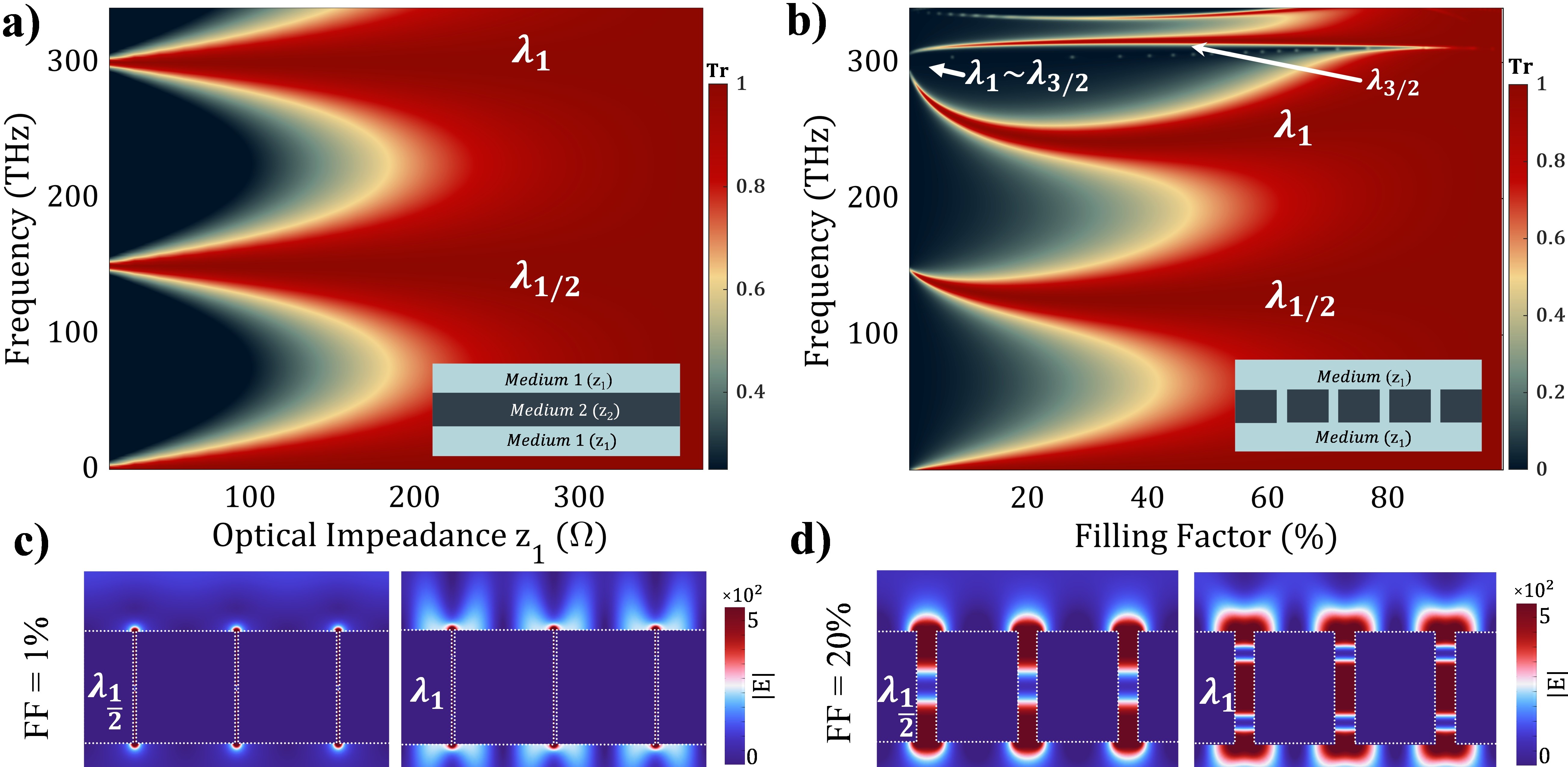}
\caption{\textbf{a)} Transmission characteristic of a dielectric FP resonator modeled as a three-layer system $z_1-z_{2}-z_1$ (see inset). \textbf{b)} Optical transmission map for a PEC grating as a function of frequency and aperture filling factor, illustrating the evolution of FP resonant bands ($\lambda_{1⁄2},\lambda_1,\lambda_{3/2}$). \textbf{c)} and \textbf{d)} Modulus of the electric field at resonance for two aperture filling factor $1 \%$ and $20 \%$, respectively.}
\label{fig:Figure 4}
\end{figure}
\section{Enhanced Transmission in a Perfect Electric Conductor}\label{sec4}
To isolate the geometric contribution behind the enhanced transmission and isolate it from the material's optical properties, we first consider the structure as an ideal PEC, modeled via Dirichlet boundary conditions ($E_\parallel=0$), eliminating all dispersive and absorptive effects. In this limit, transmission occurs solely through FP resonances, with the apertures acting as subwavelength cavities. The axial resonance condition determines the number of half standing-wave confined laterally within the slits. The grating period $H$, in contrast, dictates the diffraction onset \cite{oubeniz2023controlled, mangach2023symmetrical}. Fig~\ref{fig:Figure 3}.a and b display the FEM and FDTD transmission simulations for 1 µm and 100 nm periodicities, with a 20\% aperture size, which exhibit the three Lorentzian peaks associated with the $\lambda_{1/2},\lambda_1,\lambda_{3/2}$ resonances. The spectra are self-similar across scales, as expected for a PEC grating where frequency simply scales inversely with geometry. The highest-frequency peak $\lambda_{3/2}$ marks the transition to diffraction-dominated behavior, confirming the modal picture introduced in the previous section. The observed discrepancy in the transmission amplitude of $\lambda_{3/2}$ between FEM and FDTD arises because this resonance is located close to the diffraction threshold, where the spectral response becomes highly sensitive to solver-specific signatures and numerical formulation.\\
Importantly, the resonance positions of this aperture-based resonator agree with a rudimentary FP interpretation (Fig~\ref{fig:Figure 4}.a). By analogy with a dielectric FP resonator formed by impedance mismatch \cite{hodgson2005laser}, the grating can be modeled as an effective medium of impedance $z_\text{2}$ sandwiched between two regions (air) of impedance $z_1$. In other words, forming an aperture-based implementation of the FP resonator \cite{shen2005mechanism}. This approximation captures the main resonance behavior dictated purely by geometry. A parametric study of the aperture filling factor (Fig~\ref{fig:Figure 4}) further clarifies this behavior. Large apertures ($>50 \%$) yield near-unity transmission characteristic of a propagative regime. As the aperture decreases, spectrally Lorentzian resonances emerge and progressively narrow. This linewidth reduction reflects the transition from propagation-dominated transport to a cavity-driven, resonance-dominated regime. To ensure well-defined FP resonances, the filling factor must be sufficiently small, typically below $\sim 35 \%$. Furthermore, for very low aperture sizes, the fields become more tightly localized within the slits, as depicted in the field maps of Fig~\ref{fig:Figure 4}.c. This aperture effect causes a resonance shift due to field leakage and inter-aperture coupling. Deviations from the ideal FP behavior therefore arise from geometric confinement effects. These deviations are captured through a geometry-dependent correction factor. This is consistent with established literature,e.g., Takakura’s analysis of PEC slits \cite{takakura2001optical} and subsequent microwave experiments \cite{yang2002resonant}. Such deviations can be captured through a geometrical correction factor $\sigma(r)$, quantifying the shift from the ideal FP frequency. This factor is evaluated by dividing the reference FP resonance frequency by that exhibited in an aperture-based FP implementation, which we summarized in Appendix A.Fig~\ref{fig:Figure A9}. For example, for the aperture size we opted for in the paper (20\%), $\sigma_{20\%}\approx300/245=1.22$ for $\lambda_1$ mode. This factor encapsulates the purely geometric modification of the resonant response, forming a reference baseline for the plasmonic case discussed next.
\begin{figure}
\centering
\includegraphics[width=0.98\linewidth]{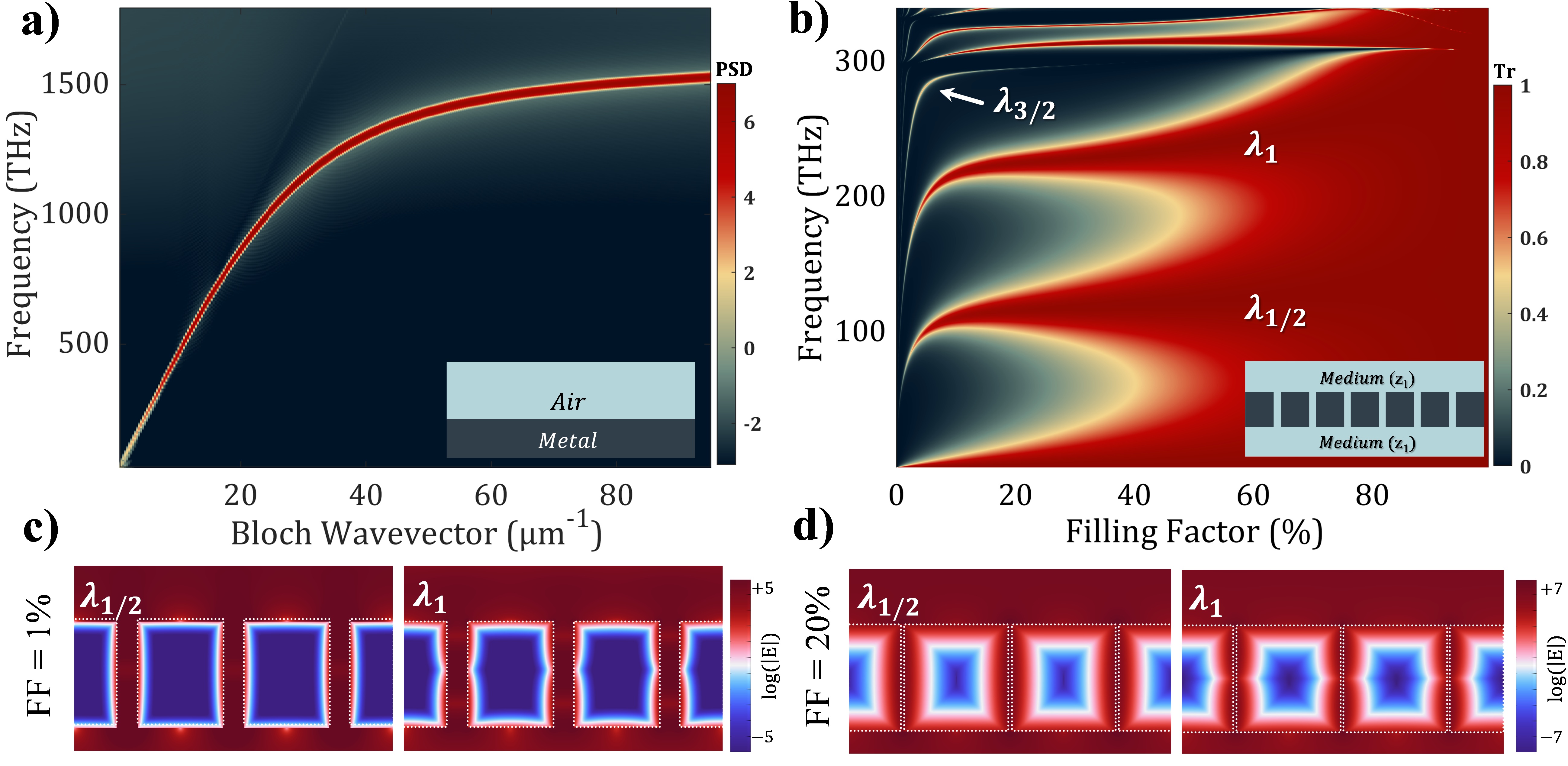}
\caption{\textbf{a)} Classic SPP dispersion characteristic, with the color map reflecting PSD. \textbf{b)} Color map of the transmission outlining the effects of filling factor versus frequency in the case of a dispersive metal with the FP resonant bands $(\lambda_{1⁄2},\lambda_1,\lambda_{3/2})$ being depicted. \textbf{c)} and \textbf{d)} Modulus of the electric field in a logarithmic distribution at resonance for two aperture filling factor $1 \%$ and $20 \%$, respectively.}
\label{fig:Figure 5}
\end{figure}
\section{Enhanced Transmission in a Dispersive Metal}\label{sec5}
In this section, we examine the transmission of electromagnetic waves through plasmonic structures while accounting for metallic dispersion, modeled by the Drude relation (parameters defined earlier). Fig~\ref{fig:Figure 5}.a shows the characteristic optical dispersion of a metal–dielectric interface, where frequency is plotted against the parallel component of the reduced Bloch vector. At low frequencies, the plasmonic response resembles that of a PEC, with the SPP curve following the light line with a similar group velocity. As frequency increases, however, the SPP branch deviates into the evanescent region and asymptotically approaches zero group velocity, capturing the transition from near-perfect conductivity to strongly dispersive plasmonic behavior.\\
Fig~\ref{fig:Figure 5}.b illustrates how aperture size influences transmission when dispersion is included. For large apertures (low impedance), the response remains PEC-like, characterized by high transmission and minimal resonance. In contrast, narrower apertures exhibit red-shifted and attenuated FP resonances due to the frequency-dependent, lossy permittivity of real metals. The stronger the field confinement within subwavelength apertures (smaller filling factors), the more pronounced the near-field interactions become at the metal–dielectric interfaces, as evidenced by the contrasting field penetration into the metal blocks shown in Fig~\ref{fig:Figure 5}.c and d. This enhanced confinement gives rise to SPP-like behavior, where the propagating wave experiences an increased effective refractive index and a corresponding reduction in group velocity. This mechanism underpins extraordinary optical transmission (EOT), where plasmon-assisted coupling enables sub-diffraction transmission. Unlike the PEC case, where resonance shifts arise solely from geometric confinement, the dispersive and absorptive response of the metal introduces additional frequency-dependent phase contributions. Both effects compound to perturb the ideal FP behavior. To account for both effects, we introduce a composite correction factor, $\sigma(r,\varepsilon)$, which captures the combined geometric and material-induced shifts.\\
\begin{figure}[h]
\centering
\includegraphics[width=1\linewidth]{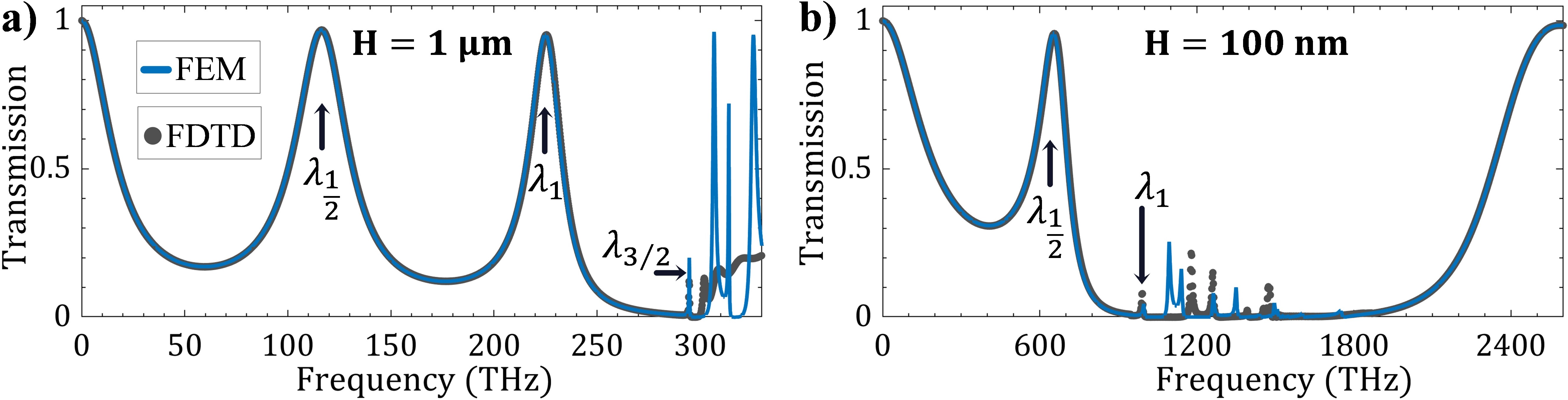}
\caption{\textbf{a)} and \textbf{b)} Transmission spectra evaluated using FEM and FDTD for two different sizes $H=1 \mu$m and $H=100$ nm, respectively, at preset filling factor of $20\%$. The $\lambda_{3/2}$ FP resonant peak has been obscured by diffraction anomalies for $H=100$ nm.}
\label{fig:Figure 6}
\end{figure}
The pivotal distinction from the PEC case is the altered transmission response under size scaling, outlined in Fig~\ref{fig:Figure 6}. For $H = 1 \mu$m, the response closely follows the PEC behavior (Fig~\ref{fig:Figure 3}.a) whereas for $H = 100$ nm, strong plasmonic coupling produces substantial red-shifts of the FP bands, from $1277$ and $2454$ THz in the PEC case (Fig~\ref{fig:Figure 3}.b) to $583$ and $987$ THz for the first and second modes, respectively, which reflects pronounced wave retardation effects. The diffraction band observed in the PEC nearly vanishes, replaced by a broad transmission plateau at higher frequencies, and the overall amplitude decreases due to ohmic losses. The FP resonance plays a central role in mediating SPP coupling. Fields during resonance are firmly confined within apertures and leak to the plasmonic material through the skin effect. This FP-SPP modal hybridization is what we have explored in section 3, showing that indeed there was major field leakage into the material (Fig~\ref{fig:Figure 2}.c). Subsequently, optical energy is transmitted or absorbed at a rate correlated to the imaginary part of the dielectric permittivity, thereby tying the transmission intensity directly to the material’s plasmonic performance and linking resonance behavior to underlying electronic properties.
\section{Plasmonic Dispersion Mapping via Fabry--P\'erot Resonance}\label{sec6}
\begin{figure}
\centering
\includegraphics[width=0.95\linewidth]{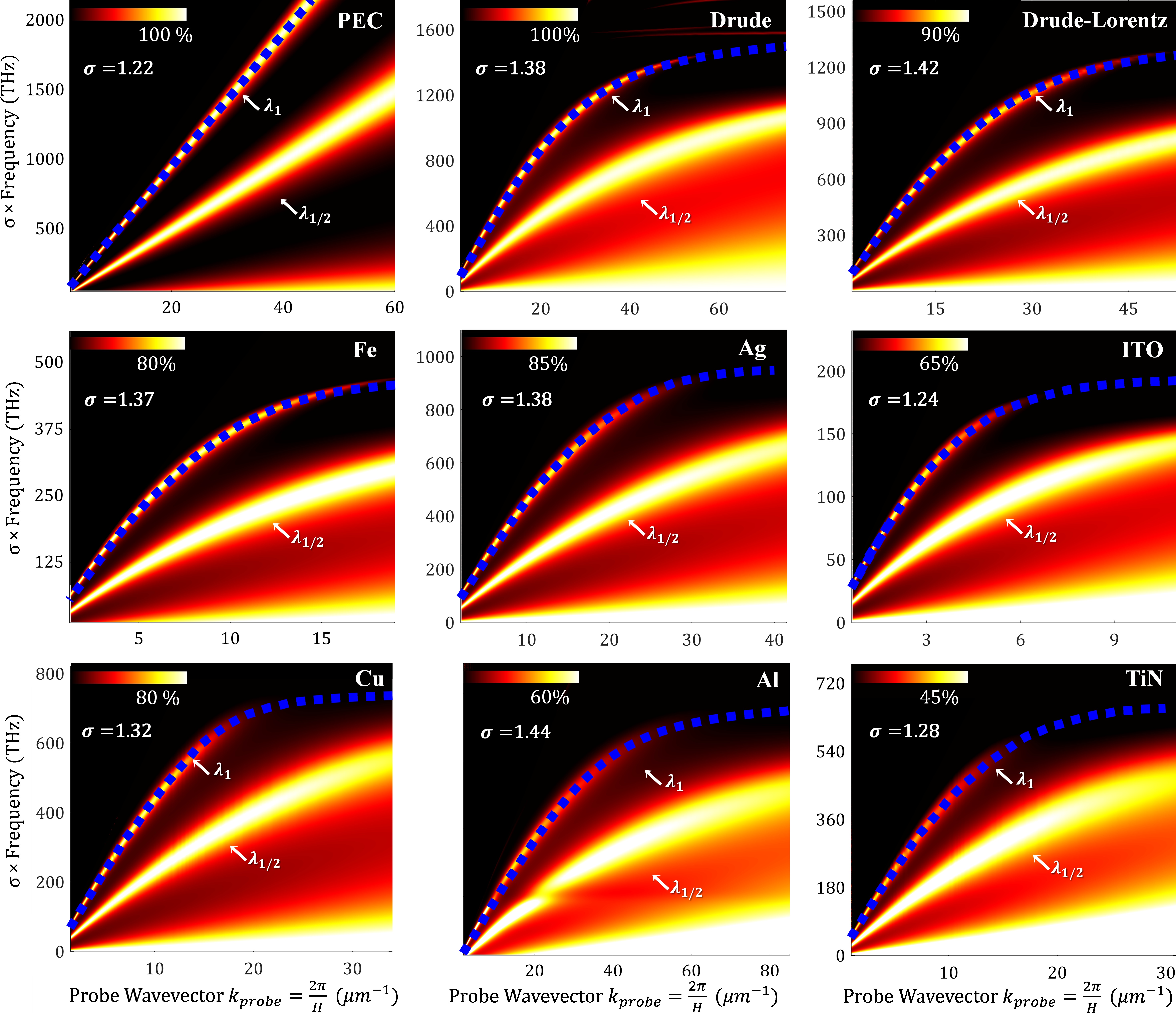}
\caption{Plasmonic dispersion mapping via FP resonance. Optical transmission maps showing resonant bands associated with FP cavity modes ($\lambda_{1/2}$, $\lambda_1$). Transmission colormap percentage scale at the top left is a proxy for plasmonic efficiency. The corrected resonance frequency $\omega_{\text{corr}}(r,\varepsilon)$ is plotted along the y-axis, while the x-axis shows the probe wavevector $k_{\mathrm{probe}}=2\pi/H$. The dashed blue curve is the SPP dispersion characteristic obtained from eigenmode analysis of the same material-air interface.}
\label{fig:Figure 7}
\end{figure}

As a proof of concept, we apply the angle-independent dispersion mapping procedure to several representative plasmonic models and experimental datasets. A PEC (non-dispersive reference), a Drude metal (explored previously), Drude-Lorentz ($\varepsilon_\infty=1$, $\omega_p=1.258\times10^{16}$ rad/s, $\gamma=7.293\times10^{13}$ rad/s, $\omega_0=1.519\times10^{12}$ rad/s), and experimental dataset parameterizations for Al \cite{mcpeak2015plasmonic}, Cu \cite{babar2015optical}, Ag and Fe \cite{palik1998handbook}, as well as representative alternative plasmonic materials for ITO \cite{naik2013alternative} and TiN \cite{beliaev2023optical}. For the mapping process, we treat each period ($H$) as a discrete momentum probe and plot resonance frequency versus \(k_{\text{probe}}\). The grating periodicity defines the dominant reciprocal-lattice vector available for SPP coupling. Consequently, the FP resonance samples the SPP branch in the vicinity of that momentum. Concretely, we define the probe momentum as the first reciprocal-lattice vector, $G=2\pi/H$, and associate each grating with $k_{\mathrm{probe}}=mG$ (here m=1). The grating period $H$ is selected individually for each material by mapping the fundamental FP resonance wavevector $k_1=2\pi/H$ onto the PEC–dielectric dispersion relation $\omega=ck$, placing the corresponding resonance frequency $\omega_1=2\pi c/H$ just below the plasmonic asymptote and thereby centering the period sweep over the most informative momentum range. Consequently, the procedure requires only approximate a priori knowledge of the material's plasmonic spectral range to define the periodicity sweep and knowledge of the surrounding dielectric permittivity to establish the reference light line.\\
The mapped SPP dispersion is recovered by correcting the FP resonance frequencies extracted from the transmission spectra. This is achieved using a dimensionless calibration factor $\sigma(r,\varepsilon)=\sigma_{geo}(r)+\Delta\sigma_{mat}(\varepsilon)$, according to: $\omega_{\text{corr}}(r,\varepsilon) \;=\; \sigma(r,\varepsilon)\;\omega_{\text{res}}(r,H)$, where \(\omega_{\rm res}\) is the resonance frequency. As previously discussed, the geometric factor $\sigma_{geo}(r)$ accounts for aperture-induced field leakage and inter-cavity coupling deviating from ideal FP behavior. The residual term $\Delta\sigma_{mat}(\varepsilon)$ encapsulates additional phase shifts introduced by the inherent optical permittivity of the material. The total correction factor $\sigma(r,\varepsilon)$ is determined through a one-time numerical calibration, and is included for each panel. First, FP resonance frequencies $\omega_{res}$ are extracted from the computed transmission spectrum. Second, at low frequencies where the SPP mode approaches the light line and material dispersion is negligible ($\Delta\sigma_{\mathrm{mat}}\approx0$), the corrected frequency $\omega_{corr}$ must coincide with the light line at the corresponding wavevector by simply multiplying the resonance frequencies $\omega_{res}$ by $ \sigma(r,\varepsilon)$. Third, the ratio $\sigma=\omega_{corr}/\omega_{res}$ is evaluated at these low-frequency calibration points to yield a single scalar per $(r,\varepsilon)$ configuration. Finally, this scalar is applied uniformly across the full spectral window to produce $\omega_{corr}$, which is then compared against the independently computed eigenmode SPP dispersion of the same material-air interface. The present procedure should therefore be interpreted as a calibrated retrieval method rather than a closed-form predictive model.\\
Fig~\ref{fig:Figure 7} presents parametric dispersion maps as a function of corrected frequency \(\omega_{\text{corr}}\) versus probe momentum $(2\pi/H)$. The transmission intensity depicted by the color scale percentage is a proxy for the material's plasmonic efficiency. For reference, the rigorous flat-interface SPP dispersion characteristic obtained using eigenmode analysis of the same materials interfacing air is overlaid with the dashed blue line. In principle, the dispersion relations can be mapped using either the fundamental ($\lambda_1$) or the half-order ($\lambda_{1/2}$) FP modes, with both yielding equivalent physical information at a given frequency. However, the $\lambda_{1/2}$ mode requires a cavity size that is half as long, making it less favorable for nanofabrication. In the PEC limit, where the permittivity is frequency-independent, the FP resonance is governed solely by the geometric FP condition, with the effective cavity-length correction arising entirely from geometric effects, in contrast to dispersive materials where material-dependent corrections are also present. When mapped against probe wavevector \(k_{\text{probe}}=2\pi/H\) for our fixed \(L/H\) scaling, appears as a linear light-line relation and provides the baseline calibration for \(\sigma(r)\). Introducing metallic dispersion leads to deviations from the light line and the emergence of asymptotic SPP behavior. These trends are consistent with the eigenfrequency dispersion models. The FP-SPP hybrid resonances curve toward the SPP dispersion and the amount of curvature depends on the real part of $\varepsilon(\omega)$, confirming that $\sigma$ must include both geometric and dielectric contributions. Experimental datasets display distinct signatures: low-loss materials produce sharp, high intensity FP peaks whose corrected positions closely track the eigenmode SPP. The lossy or interband-rich materials show broader resonance bands with lower optical transmission amplitudes, particularly at higher frequencies. Reconstruction quality degrades for strongly lossy materials (e.g., Al, ITO and TiN) where resonances broaden and amplitude is suppressed. Finally, the maps expose a practical constraint in the case of strongly lossy materials, for which the usable spectral window must be restricted to regions with resolvable bands.\\
The geometry underlying this method, namely, an array of discrete metallic ridges separated by subwavelength gaps in a symmetric dielectric environment, has been extensively studied in nanophotonics \cite{garcia2010light, murugan2024recent}. A common fabrication route begins by defining the slit pattern in a metallic layer by focused ion beam (FIB) milling, after which the patterned film is coated with the dielectric cladding \cite{guillaumee2010observation, mahani2024plasmonic}. Alternatively, the ridge array can be patterned directly on a dielectric or metal-coated substrate by electron-beam lithography (EBL) with lift-off \cite{sreekanth2014improved}, EBL followed by top-down etching \cite{mukherjee2025spectroscopic}, laser interference lithography with dry etching \cite{arriola2012fabrication}, EBL with plasma etching and electroplating \cite{yue2024highly}, and block-copolymer self-assembly with image-reversal etching \cite{jung2010fabrication}. Once the ridge array has been defined, the encapsulation strategy is selected according to the desired dielectric material and the structural requirements of the device. In the symmetric embedding configuration targeted here, the same dielectric constitutes the substrate, superstrate, and slit infill. The straightforward route is polymer spin-coating (PMMA, PDMS, BCB, SU-8, or Cytop), in which the liquid encapsulant fills the subwavelength inter-ridge gaps by capillary action while the excess material simultaneously forms the superstrate, at a single ambient condition step that requires neither vacuum equipment nor alignment \cite{murugan2024recent, guillaumee2010observation, huang2017characterization}. When the desired dielectric cannot be deposited from solution, vacuum-based conformal deposition provides an alternative. Atomic layer deposition (ALD) of SiO$_2$ has been shown to conformally fill metallic grating slits with aspect ratios exceeding 2 \cite{kang2013enhanced}, and DC magnetron reactive sputtering has been used to encapsulate gold ridge gratings with TiO$_2$ \cite{lamy2017broadband}. PECVD has been used to embed silver nanostructures within an ultrathin amorphous silicon layer, demonstrating conformal embedding of isolated sub-20-nm height features \cite{ahmed2025machine}. Non-conformal growth morphology has also been reported in thicker device stacks on nanotextured or imprinted metallic back reflectors, where deposition shadowing \cite{cao2020light} and planarizing growth \cite{paetzold2011plasmonic} progressively smooth or invert the original substrate profile. Appendix B, Fig.~\ref{fig:Figure A10}, presents dispersion maps for the same Lorentz--Drude material interfaced with PDMS ($\varepsilon=2.56$) and Si ($\varepsilon=12$). In both cases, the simulations use the same dielectric material for the substrate, superstrate, and slit infill. The Si case is included to show that the dispersion-mapping procedure remains valid across the dielectric-constant range relevant to realistic embedding media, rather than to suggest Si as a specific practical implementation. Accordingly, the spin-coated polymer configuration remains the primary experimentally realizable implementation considered in this work.\\
The method is inherently non-broadband: each grating periodicity samples a single in-plane momentum state, and the full dispersion curve is assembled by systematic period variation across a set of structures. Precise periodicity and aperture dimensions remain important, and the influence of realistic aperture wall roughness on the FP modal response is examined in the following section. The analysis demonstrates that the measurement modality retains its physical integrity under realistic fabrication conditions. In many technologically relevant contexts, the discrete probing strategy is not a limitation but a direct advantage. Platforms operating at fixed spectral lines, telecom devices near 1.55 $\mu$m, mid-infrared QCL-based sensing systems, or single-wavelength plasmon-enhanced detectors require no broadband reconstruction, and a single appropriately designed grating suffices to evaluate coupling efficiency and modal dispersion at the target frequency. Where full dispersion reconstruction is required, systematic period variation enables wafer-scale mapping and process quality control. The method also carries a physically distinctive sensitivity not directly achieved in conventional planar metrology. Ellipsometry and SPR spectroscopy interrogate flat, extended interfaces and do not capture, in the same way, the optical response of the vertical sidewall surfaces produced by subwavelength patterning. The FP cavity mode is bounded precisely by these walls, and its hybridization frequency is governed by their dielectric response. Process-induced modifications localized at these non-planar surfaces, such as gallium implantation from FIB milling, carbonaceous deposition from EBL, or acid residues from wet etching, are therefore encoded directly in the cavity eigenfrequency. The technique is further spatially discriminating: local dimensional drift or inter-device process variation produces measurable resonance shifts that are washed out in area-averaged bulk spectra, making the method naturally suited for inline inspection and failure analysis in plasmonic device manufacturing.
\begin{figure}
\centering
\includegraphics[width=0.95\linewidth]{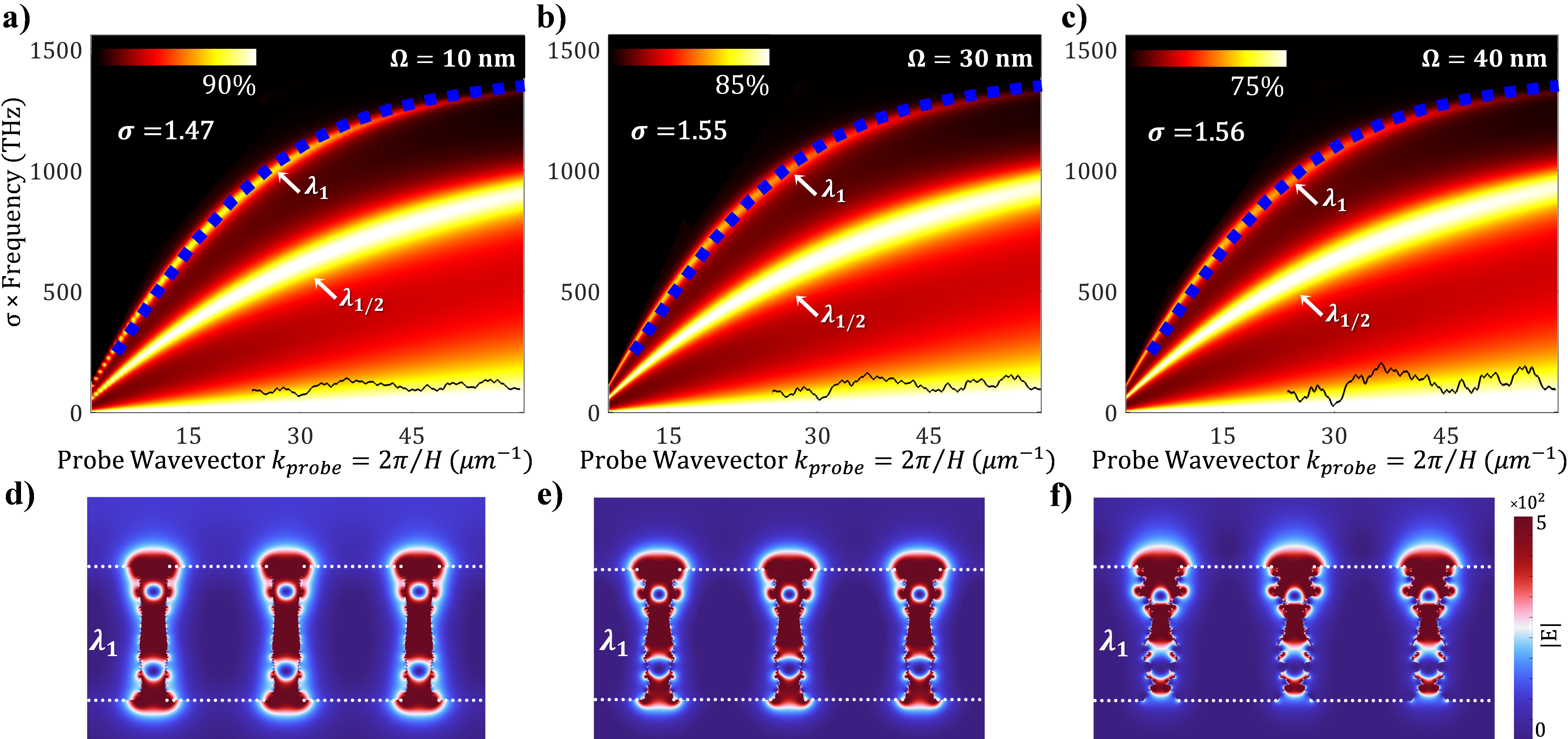}
\caption{\textbf{a-c)} Plasmonic dispersion mapping for a Drude–Lorentz metal via FP resonance under increasing cavity-wall roughness. The zigzag lines indicate the rough cavity boundaries. The corrected resonance frequency $\omega_{\text{corr}}(r,\varepsilon)$ is plotted as a function of the probe wavevector $k_{\mathrm{probe}}=2\pi/H$. The transmission colormap percentage scale serves as a proxy for plasmonic efficiency. The dashed blue curve denotes the SPP eigenmode dispersion of the same material–air interface. \textbf{d-f)} Modulus of the electric field at the $\lambda_1$ resonance for roughness amplitudes $\Omega = 20$ nm, $30$ nm, and $40$ nm, corresponding to apertures with filling factors of approximately $20\%$, $26\%$, and $27.5\%$, respectively. The white dotted lines mark the geometrical boundaries.}
\label{fig:Figure 8}
\end{figure}
\section{Influence of Fabrication-Induced Roughness}\label{sec7}
To assess the tolerance of the dispersion-mapping procedure to fabrication-related imperfections, the Drude-Lorentz material defined in the previous section was re-examined for stochastically rough aperture walls. Fig~\ref{fig:Figure 8} a)-c) show the dispersion maps for this material under the three roughness configurations, corresponding to roughness amplitudes $\Omega=$ $20$ nm, $30$ nm and $40$ nm, for a nominal aperture size of $200$ nm. The primary observation is a clear decoupling between spectral position and transmitted amplitude: the Lorentzian resonance peaks associated with the $\lambda_{1/2}$ and $\lambda_1$ FP modes retain their spectral positions across all three configurations, with no measurable linewidth broadening and any residual frequency shift remaining within the correction range of the calibration factor. The perturbation to the walls changes the effective filling factor of the aperture to approximately $20\%$, $26\%$, and $27.5\%$ for the three configurations, respectively, which accounts for the change in the correction factor $\sigma(r,\varepsilon)$. The transmitted amplitude, by contrast, decreases monotonically with increasing $\Omega$, while the dispersion maps remain spectrally faithful to the smooth-wall reference throughout.\\
The electric-field modulus maps shown in Fig.~\ref{fig:Figure 8} d)-e) elucidate the mechanisms underlying both observations. Despite the local distortion introduced by the rough boundaries, the field distribution in the rough-wall case retains the recognizable modal topology of the smooth-wall $\lambda_1$ eigenmode: the longitudinal standing-wave pattern and antinodal structure are preserved, with roughness producing local intensity redistribution rather than a reorganization of the global cavity mode. This persistence of modal character is consistent with the absence of significant eigenfrequency shifts. The FP resonance condition is governed by phase accumulation over the full cavity length, a spatially extended quantity, and is therefore relatively insensitive to localized boundary perturbations that do not alter the global modal topology. The enhanced metallic absorption responsible for the reduced transmission amplitude has a clear geometric origin. The field maps reveal localized intensity enhancement at both wall protrusions and recesses, increasing the overlap of the electromagnetic field with the finite skin depth of the metal and thereby enhancing ohmic dissipation. Because this effect becomes more pronounced with increasing perturbation amplitude $\Omega$, it directly accounts for the observed monotonic decrease in the transmission peak.\\
This behavior stands in contrast to conventional prism-coupled or grating-coupled SPP spectroscopy, in which the plasmonic field is concentrated at the metal-dielectric interface, precisely the region where fabrication roughness acts. In that configuration, surface perturbations can directly modify the local SPP propagation condition, producing measurable resonance broadening and spectral displacement. In the FP--SPP architecture, by contrast, the resonance condition is set by a cavity mode localized within the aperture volume and coupled to the SPP response through modal hybridization. The probing observable is therefore less directly exposed to interface-bound perturbations that would otherwise corrupt an angle-resolved measurement. The field maps in Fig.~\ref{fig:Figure 8} make this point explicit: despite the rough-wall distortion, the mode remains recognizably the same object as its smooth-wall counterpart. The residual influence of roughness on the mapping procedure enters primarily through the effective filling factor. The stochastic boundary displacements alters aperture size relative to the nominal geometry, modifying the coupling amplitude and the effective resonator volume. This geometric effect is largely absorbed by the $\sigma(r,\varepsilon)$ calibration factor without compromising the spectral positions of the mapped resonances. The dispersion reconstruction accordingly remains accurate across all three roughness amplitudes, confirming the practical robustness of the proposed mapping scheme under conditions representative of state-of-the-art nanofabrication.
\section{Conclusion}
We have presented a computational framework for mapping plasmonic dispersion relations from transmission through metallic gratings configured as subwavelength FP resonators. The method exploits hybridization between FP cavity modes and SPPs, so that resonance shifts encode both geometric confinement and material dispersion. A key result is that the dispersion correction can be obtained from a simple calibration of the low-frequency linear regime, aligned to the light line, with only coarse prior knowledge of the plasmonic asymptote needed to set the sweep range. Once this correction is established, the flat-interface SPP dispersion can be reconstructed from transmission resonances without angular scanning.\\
We demonstrated the approach on PEC, Drude, and Drude-Lorentz models, as well as experimental data for Fe, Ag, Al, Cu, ITO, and TiN, showing its generality and its sensitivity to losses, interband effects, and confinement. For high-loss materials, the usable spectral window is reduced by damping. Beyond dispersion reconstruction, the framework suggests a broader role for FP-assisted grating probes as a compact plasmonic metrology tool. Fabrication imperfections were also examined by introducing stochastic roughness along the metallic boundaries. The mapping remained robust across the three roughness amplitudes, with no appreciable broadening of the FP modes, although transmission decreased due to enhanced ohmic losses.\\
\appendix{\textbf{Appendix A: Details on Geometrical Correction Factor}}
\begin{figure}[h]
\centering
\includegraphics[width=0.5\linewidth]{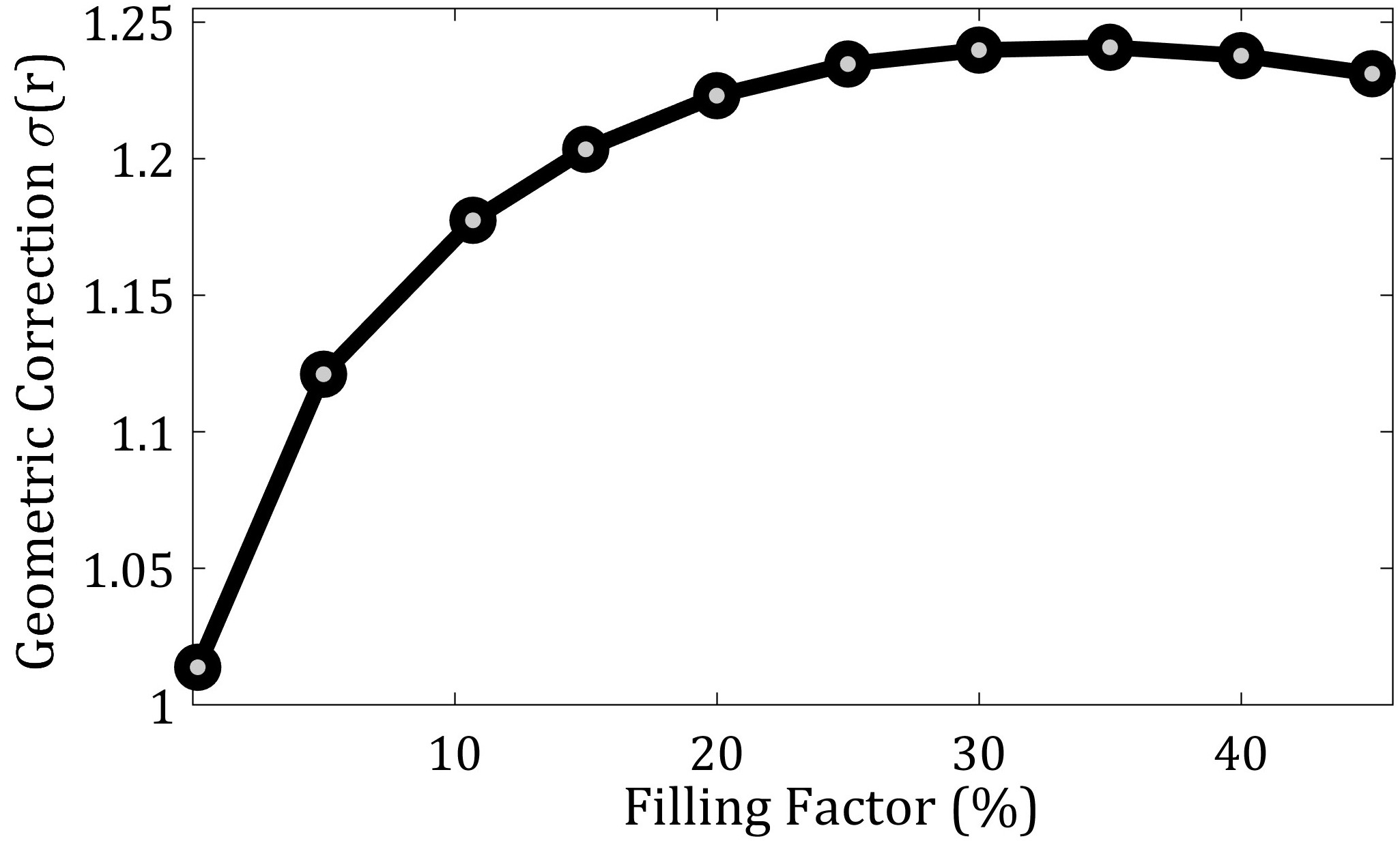}
\caption{Effect of aperture filling factor on the $\lambda_1$ band for a PEC metal. The geometrical correction factor $\sigma(r)=\frac{f(\text{dielectric})}{f(\text{aperture})}$ characterizes the frequency deviation from a reference FP induced by aperture confinement.}
\label{fig:Figure A9}
\end{figure}
\appendix{\textbf{Appendix B: SPP Mapping for different dielectric environments}}
\begin{figure}[h]
\centering
\includegraphics[width=0.9\linewidth]{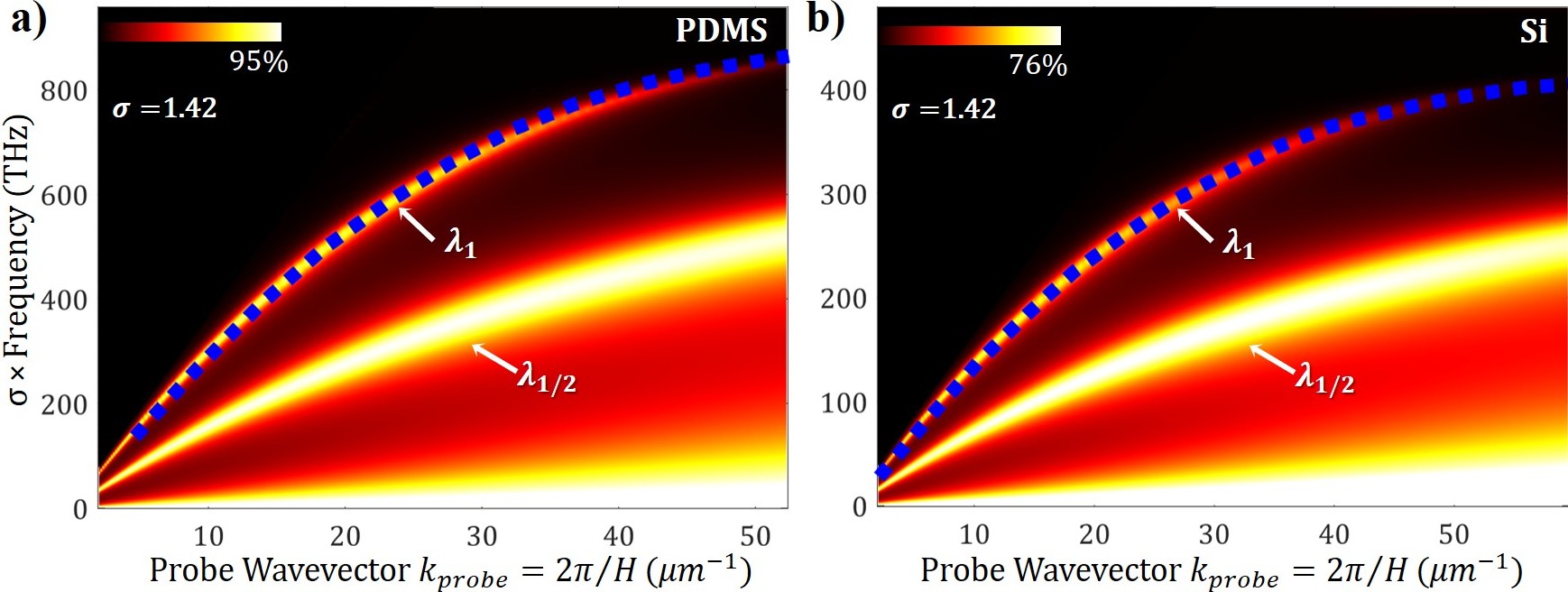}
\caption{ \textbf{a)} and \textbf{b)} Plasmonic dispersion mapping via FP resonance for the same Drude-Lorentz metal interfacing PDMS ($\varepsilon=2.56$) and Silicon ($\varepsilon=12$), respectively. Optical transmission maps showing resonant bands associated with FP cavity modes ($\lambda_{1/2}$, $\lambda_1$). Blue dashed curves are the SPP Eigenmode signatures.}
\label{fig:Figure A10}
\end{figure}
\ack{The authors thank Prof. Philippe Lalanne for making his team's implementation of the auxiliary field expansion weak form available under an open-source license.}
\roles{All authors have read and contributed to the writing of the paper.}
\data{The data and numerical models that support the findings of this study are available from the corresponding author upon reasonable request.}
\bibliographystyle{iopart-num}
\bibliography{sample.bib}
\end{document}